\documentclass[aps,showpacs]{revtex4}

\usepackage{graphicx}

\def\bm#1{{\mbox{\boldmath $#1$}}}
\def\omegaBP{\omega_{\mathrm BP}}
\def\rhoc{\rho_{\mathrm c}}

\newcommand{\be}{\begin{equation}}
\newcommand{\ee}{\end{equation}}

\begin{document}
  
\title{Vibrations in glasses and Euclidean Random Matrix theory}

\author{T.~S.~Grigera}
\author{V.~Mart\'\i{}n-Mayor} 
\author{G.~Parisi}

\affiliation{Dipartimento di Fisica, Universit\`a di Roma ``La
Sapienza'', P.le Aldo Moro 2, 00185 Roma, Italy INFN sezione di
Roma - INFM unit\`a di Roma}

\author{P.~Verrocchio}
\affiliation{Dipartimento di Fisica, Universit\`a
di Trento, Via Sommarive, 14, 38050 Povo, Trento, Italy INFM unit\`a
di Trento}

\date{October 18, 2001}

\begin{abstract}
We study numerically and analytically a simple off-lattice model of
scalar harmonic vibrations by means of Euclidean random matrix theory.
Since the spectrum of this model shares the most puzzling spectral
features with the high-frequency domain of glasses (non-Rayleigh
broadening of the Brillouin peak, boson peak and secondary peak), the
Euclidean random matrix theory provide a single and fairly simple
theoretical framework to their explanation.
\end{abstract}
\pacs{PACS}
\maketitle

\section{Introduction}

The high-frequency ($\omega > 1Thz$) region in vibrational spectra of
amorphous systems is related to density fluctuations whose size is
comparable to the typical distance between particles.  Whereas in
ordered systems those excitations (phonons) persist up to momenta of
about the Debye momentum, the fate of excitations of microscopical
size in disordered systems is still a quite puzzling issue, both from
the theoretical and the experimental point of view.

Recent high-resolution inelastic X-ray scattering (IXS) and neutron
scattering techniques have made accessible to experiment this
region~\cite{Buchenau86,silica,glycerol,Foret96,Fioretto99,Masciovecchio98,
Monaco98,Ruocco99,Sokolov99,Matic01,Sokolov95,Tao91,Engberg99,Wischnewski98}
where the exchanged external momentum $p$ is comparable to $p_0$,
namely the momentum where the static structure factor has its first
maximum. A related issue is the influence of the high-frequency
dynamics of glasses on specific heat experiments in the
$1\,$K--$100\,$K range.  A number of facts have emerged from the
experiments:

\begin{enumerate}
\item The dynamic structure factor $S(p,\omega)$ has a {\bf
Brillouin-like peak} for momenta up to $p/p_0 \sim 0.5$.  This
inelastic peak is due to the interaction of the external photon (or
neutron) with some excitation of the system.  A very controversial
issue is the propagating nature of those
excitations~\cite{SilicaDebate}.  Furthermore a {\bf secondary peak}
at frequencies smaller than the Brillouin one develops for larger
momenta~\cite{talks,Horbach01}, becoming dominant for $p/p_0 \sim
0.5$.

\item The Brillouin peak width $\Gamma$ dependence on the momentum
$p$ has been described by means of the following scaling law:
\begin{equation}
\Gamma \propto p^{\alpha}.
\label{BROAD}
\end{equation}
The broadening must generally be ascribed to the interaction among the
excitations and the disorder of the system.  However the details are
still unexplained, two remarkable facts indeed arising.  First, there
seems to be quite a general agreement about the fact that $\Gamma$ is
not affected by changes in the temperature, ruling out the
hydrodynamic explanation for the broadening. Second, at small enough
momenta (e.g. light scattering) $\alpha$ is undoubtedly $\sim 2$,
ruling out also Rayleigh scattering.

\item Beyond the well-known anomalous linear dependence on $T$ at the
lowest temperatures of the specific heat (reproduced by the two-level
model~\cite{twolevel}), another deviation from the Debye $T^3$ law at
slightly higher temperatures is found. The specific heat divided by
$T^3$ shows a peak near $10\,$K~\cite{pohl81}.  This peak points to
the existence of an excess of vibrational states at frequencies
$\omega \sim 1\,$THz, which shows as a peak in the plot of
$g(\omega)/\omega^2$ ($g(\omega)$ is the vibrational density of
states, as obtained by Raman or inelastic neutron scattering). Since
the scattering intensity at the peak scales in temperature with Bose
statistics, the peak has become known as the {\bf Boson peak} (BP).
Two features of the BP should be remarked. First, the peak frequency
$\omegaBP$ is several times smaller than any natural frequency scale,
like the Debye frequency or the band edge. Moreover, where good data
for the dispersion relation $\omega(p)$ (determined from the position
of the Brillouin peak of the dynamic structure factor, which probes
only longitudinal modes) are available, it has been checked that
$\omega(p)$, is still a linear function of the momentum $p$ at
$\omegaBP$~\cite{glycerol,Matic01,silica,Fioretto99}.  Second,
although less experimental data are available, it seems to be the rule
that $\omegaBP$ shifts to lower frequencies on
heating~\cite{Sokolov95,Tao91,Engberg99} (except for
silica~\cite{Sokolov95,Wischnewski98}). Yet, in silica the BP
evolution upon increasing the density has been studied experimentally
\cite{SILICADENSITYEXP} and in simulations \cite{SILICADENSITYNUM}.
In close agreement with our most recent theoretical
results~\cite{BOSONPEAK}, it was found that $\omegaBP$ shifts to
larger frequencies and the BP loses intensity when the density grows.
The BP has also been identified with the secondary peak in the dynamic
structure factor at high exchanged momentum, both experimentally
\cite{talks} and in simulations \cite{Horbach01}.

\end{enumerate} 


A number of basic insights on the spectral properties of glasses have
been obtained by means of molecular dynamics simulations
\cite{Horbach99,Taraskin99,Ruocco00,GoMa,Grigera01,Schirmacher98,vanHove,
Kantelhardt01} on systems such as argon, silica and water.  Among
others, let us point out the fact that, in the glass phase, the high
frequency dynamics can be understood in the framework of the {\em
harmonic approximation} (see however the work on
soft-potentials~\cite{soft-potentials} for a dissenting view).


We still lack an universally agreed upon theoretical interpretation of
the propagation of {\em phonons} in topologically disordered systems.
The BP has been variously interpreted as arising from mixing of
longitudinal and transverse modes
\cite{Horbach01,Sampoli97,Ruocco00,Matic01}, hybridization of optical
and acoustic modes \cite{Taraskin99}, a combination of level repulsion
from disorder and van Hove singularities \cite{vanHove}, an associated
mechanical instability \cite{Grigera01,GoMa}, the presence of a
Ioffe-Regel crossover at $\omegaBP$ \cite{Foret96,Parshin01}, or from
the scattering of sound-waves with localized anharmonic
vibrations\cite{soft-potentials,Wischnewski98}.

Here we would like to describe a theoretical approach relying on the
study of the statistical properties of Euclidean Random
Matrices~\cite{MePaZe} (ERM) providing a coherent framework which
explains the broadening of the Brillouin peak and the origin both of
the Boson and of the secondary peak.

The organization of the paper is as follows. Sec.~\ref{REAL} defines
our model and gives technical some details on the perturbative
solution and resummation. Next, sections~\ref{BRILL} and~\ref{BOSON}
discuss the more general theoretical results, while in
sec.~\ref{NEARLY-GAUSSIAN} a special interparticle potential is
studied, comparing theoretical and simulation
results. Sec.~\ref{CONCLUSIONS} summarizes our conclusions.

\section{Real systems and ERM theory}\label{REAL}

We are considering the spectral properties of a system whose particles
oscillate around disordered positions.  For the sake of simplicity we
shall consider only the case the {\em scalar} case, where all the
displacements are collinear.  In such a system the spring constants
are deterministic function, depending only on the distances of the
particles rest positions, the Hamiltonian being:
\begin{equation}
H = \frac{1}{2} \sum_{i,j} \phi_i M_{i,j} \phi_j
\end{equation}
where $M$ is a matrix of the form:
\begin{equation}
M_{ij} \equiv \delta_{ij} \sum_{k=1}^N f( x_i - x_k) 
- f( x_i -x_j)\quad ,
\quad i,j=1,2\ldots,N\ .
\label{EUCLIDEA}
\end{equation}
If we think of the glass as a disordered harmonic solid, the matrix
$M$ is nothing but the Hessian matrix of the system where the function
$f(r)$ is the second derivative of the pair potential.

Hence the disorder in the interactions is due to topological reasons,
namely the disordered position of particles.  As in those
topologically disordered systems one cannot split the interactions
(spring constants) in an ordered part plus a disorder-dependent
correction, a new theoretical framework involving the so-called
Euclidean Random Matrices had to be introduced~\cite{MePaZe}.  The
quantity to be directly compared with scattering experiments is the
dynamic structure factor $S(p,\omega)$, which at one-excitation level
can be computed by means of~\cite{Martin-Mayor01}:
\begin{equation}
S^{(1)}(p,\omega)=-\frac{2 \: K_B T \: p^2}{\omega \pi}\lim_{\eta\to
0^+}{\mathrm {Im}}\ G(p,\omega^2+{\mathrm i}\eta)\,.
\end{equation}
having introduced the resolvent:
\begin{equation}
 G(p,z) \equiv \frac{1}{N} \sum_{jk} \overline{ \exp[i
 \mbox{\boldmath$p$} \cdot (
{\mbox{\boldmath$x$}}^{\mathrm eq}_j-
{\mbox{\boldmath$x$}}^{\mathrm eq}_k ) ] [(z-M)^{-1}]_{jk} }
\label{GDIPI}
\end{equation}

The overline has the meaning of average over the disordered
equilibrium positions. As a matter of fact, we are assuming that the
vibrational spectra of amorphous systems are {\em self-averaging}
quantity, hence they should not depend on the given configuration of
disorder.

A straightforward method to compute the resolvent is to write it as
the sum of a geometric series, whose $R$th element is:
\begin{eqnarray}
M^R(p)  &=& \frac{1}{N} \sum_{k_0,k_1 \ldots k_R} 
\: e^{ip x_{k_0}} \left( \delta_{k_0, k_1} 
\sum_{z_1} f(x_{k_0} - x_{z_1}) - f(x_{k_0} -x_{k_1}) \right) \dots \nonumber 
\\ &\dots& \left(\delta_{k_{R-1}, k_R} 
\sum_{z_R} f(x_{k_{R-1}} - x_{z_N}) - f(x_{k_{R-1}}-x_{k_R})\right )
e^{-ip x_{k_R}}
\label{GEO}
\end{eqnarray}
Letting $\hat f(p)$ be the Fourier transform of $f(r)$, without any
loss of generality the resolvent can be written:
\begin{eqnarray}
G(p,z)&=&\frac{1}{z-\epsilon(p)-\Sigma(p,z)}\,,\\
\epsilon(p)&=&\rho[\hat f(0)-\hat f(p)]\,
\label{LAMBDADIP}.
\end{eqnarray}
where the self-energy $\Sigma$ describes the interactions between the
phonons and the disorder.

The simplest case arises when all the particles are randomly placed
without any correlation. In that situation, using
$P[{x}]=\left(1/V\right)^N$ as the probability distribution of
quenched variables, it is possible to build a perturbative approach
where the leading term is exact when the density $\rho$ in infinite
and the next-to leading terms are proportional to powers of
$1/\rho$~\cite{Martin-Mayor01}.  Furthermore by means of well-known
resummation techniques, the $1/\rho$ correction has been exploited in
order to obtain an integral equation for the
self-energy~\cite{Grigera01}:
\begin{equation}
\Sigma(p,z)=\frac{1}{\rho}\int\frac{d^D q}{(2\pi)^D} 
\frac{ \left[\rho\left(\hat f(\mbox{\boldmath$q$})-\hat f(\mbox{\boldmath$p$}-
\mbox{\boldmath$q$})\right)\right]^2}
{z-\epsilon(q)-\Sigma(q,z)}\,.
\label{CACTUS}
\end{equation}
whose solution takes into account a given class of terms of any order.
However, the quenched position in glasses (and amorphous systems in
general) are highly correlated.  Let us sketch a simple approximation
dealing with the correlated case.  Let $g^{(R)}(x_1, \dots, x_R)$ be
the $R$-points correlation function related to an arbitrary
probability distribution $P[{x}]$:
\begin{equation} 
g(y_1, \dots, y_R) \equiv \sum^N_{j_1,\dots,j_R}
\overline{\delta(x_{j_1}-y_1) \dots \delta (x_{j_R}-y_R)}
\end{equation} 
Hence, the average on the position of the particles is:
\begin{equation}
\overline{M^R(p)} = \frac{1}{V^{R}} \int \prod^{R}_i d^d x_i
g^{(R)}(x_1 \dots x_{R}) M^R(p)
\label{RANDOM2}
\end{equation}

Although the computation using the full correlation function would be
exceedingly difficult, some progress can be made by using the so
called superposition approximation:
\begin{equation}
g(x_1 \dots x_{R+1}) = g(x_1-x_2) g(x_2-x_3) \dots g(x_R-x_{R+1})
\label{SUPER}
\end{equation}
where the pair correlation function is used to take into account the
correlation of the position of the particles. The superposition
approximation can be embedded in our calculation if we make
the substitution~\cite{Martin-Mayor01}:
\begin{equation}
 f(r) \to \ g(r)f(r)
\label{FUTURO}
\end{equation}
This is rather important, because, for typical applications, the
function $f$, being badly divergent at short distances, does not have
a Fourier transform. On the other hand, the function $g(r)$ typically
tends to zero at the origin exponentially, thus taking care of the
algebraic divergence of $f(r)$.

Let us point out that in the one-excitation approximation the
following relationship between the DOS and the dynamic structure
factor holds~\cite{Martin-Mayor01}:
\begin{equation}
g(\omega)=
\frac{\omega^2}{k_{\mathrm B} T p^2}
S^{(1)}(p\to\infty,\omega)
\label{INFINITO}
\end{equation}
As at very high frequencies the one-excitation approximation does not
hold, and many-excitations contributions should be taken into account,
the reliability of Eq.~(\ref{INFINITO}) in describing real systems is
a very interesting matter.

\section{Brillouin Peak} \label{BRILL}

The following connections between the main features of the dynamical
structure factor and the self-energy are established:
\begin{itemize}
\item
The 'bare' dispersion relation $\epsilon(p)$, which would give the
position of the peak in the elastic medium limit, is renormalized by
the real part of the self-energy $\Sigma'(p,z)$.  This gives
$\omega^{renorm}(p)$, the position of the maximum of the structure
factor in the frequency domain.  Let us note that $\omega^{renorm}(p)$
is certainly linear for small $p$, as expected.
\item The imaginary part $\Sigma''(p,z)$ 
computed at the position of the peak $\omega=\omega^{renorm}(p)$
gives the width, $\Gamma(p)$, of the $S^{(1)}(p,\omega)$ by means of:
\begin{equation}
\Sigma''(p,\omega^{renorm}(p)) = \omega^{renorm}(p) \Gamma(p)
\label{GAMMA}
\end{equation}
\end{itemize}
	
Here we want to show that eq.~(\ref{CACTUS}) provides a model-independent
derivation of the exponent $\alpha$ in the scaling law~(\ref{BROAD}). 
Indeed, the large $q$ contribution to the imaginary part of the integral 
in equation (\ref{CACTUS}) is, because of (\ref{INFINITO})
\begin{equation}
\Sigma''_0(p,z)= -\pi \rho g_\lambda(\lambda) \int\frac{d^D q}{(2\pi)^D}
\!\!  \left(\hat f(\mbox{\boldmath$q$})-\hat f(\mbox{\boldmath$p$}-
\mbox{\boldmath$q$})\right)^2.
\label{sigmazero}
\end{equation}
where $g_\lambda(\lambda)$ is the density of states in the domain of
eigenvalues ($\lambda=\omega^2,
g_\lambda(\omega^2)=\frac{g(\omega)}{2\omega}$).  If the spectrum is
Debye-like we have $g_\lambda(\lambda) \propto \lambda^{0.5}$, and it is
straightforward to show that~(\ref{sigmazero}) is proportional to
$\omega^{renorm}(p) \, p^2$. Then the relation~(\ref{GAMMA}) implies
the scaling:
\begin{equation}
\Gamma_0(p) \propto p^2 
\end{equation}
irrespective of the function $f(r)$.  Clearly this is only the large
$q$ contribution to the integral, but it has been shown that it indeed 
controls the peak width at small $p$~\cite{Grigera01}.
Hence the ERM theory yields the correct asymptotic behavior at very
low momenta of the broadening of the peak. 

\section{The Boson Peak}\label{BOSON}

Exploiting the relation~(\ref{INFINITO}), it is possible to obtain
from~(\ref{CACTUS}) an integral equation even for the DOS.
As a matter of fact, defining ${\cal G}(z) = G(p=\infty,z)$, the DOS turns
out to be
\begin{equation}
g(\omega) = - {2\omega \over \pi} \, \mathrm{Im}\, {\cal G} (\omega^2
+ i 0^+).
\end{equation} 
$\cal G$ being the solution of the following integral equation:
\begin{equation}
\frac{1}{\rho{\cal G}(z)}=\frac{z}{\rho}-\hat f(0) - A {\cal G}(z) -\int
\!\!\! \frac{d^3q}{(2\pi)^3}\hat f^2(\mbox{\boldmath$q$})
G(\mbox{\boldmath$q$},z),
\label{DENSIDADDEESTADOS}
\end{equation} 
where $A= (2\pi)^{-3}\int \!\!\hat f^2(\mbox{\boldmath$q$})\,d^3q$.
With this equation, one needs to know the resolvent at all $q$ to
obtain the DOS, due to the last term in the r.h.s. This can be done by
solving numerically the self-consistent equation of
ref.~\cite{Grigera01}, but here we perform an approximate analysis,
which is more illuminating. The crudest approximation is to neglect
this term, in which case Eq.~\ref{DENSIDADDEESTADOS} is quadratic in
${\cal G}$, and one easily finds a semicircular DOS, with center at
$\omega^2=\rho \hat f(0)$ and radius $2\sqrt{\rho A}$. This spectrum
is the glass analogue of a van-Hove singularity: indeed, when $\rho\to
\infty$, the spectrum is made of plane waves, with dispersion relation
$\omega^2(p)= \rho(\hat f(0)-\hat f(p))$,~\cite{Martin-Mayor01,MePaZe}
(a continuous elastic medium). $\omega(p)$ saturates for large $p$ at
$\omega^2=\rho\hat f(0)$, yielding an enormous pile-up of states which
causes the DOS to be concentrated at this value~\cite{MePaZe}. At
finite $\rho$, density fluctuations of the $\bm{x}^{\mathrm{eq}}$ are
present which act as a perturbation that splits this degeneracy, and
yields the semicircular part of the spectrum at high frequency.  But
the semicircular spectrum misses the Debye part, and a better
approximation is needed.  So we substitute $G$ in the last term of the
r.h.s. by the resolvent of the continuum elastic medium $G_0(z,p) = (
z - \omega^2(p))^{-1}$.  This is reasonable because the $f^2(q)$
factor makes low momenta dominate the integral, and due to
translational invariance $G(z,p) \approx G_0(z,p)$ in this
region~\cite{Martin-Mayor01}. We shall be looking at small $\omega$,
so to a good approximation
\begin{equation}
\int \!\! {d^3q \over (2\pi)^3} \, \hat f^2(q) G_0(q,z) \approx
- {1\over\rho} B - i {\rho \hat f^2(0) \over 4\pi c^3} \omega,
\end{equation}
where the sound velocity is $c=\sqrt{\rho \hat f''(0)/2}$ and
$B>0$. Then Eq.~(\ref{DENSIDADDEESTADOS}) is again quadratic in ${\cal
G}$, and can be solved to give
\begin{equation}
{\cal G}(\omega+i0^+) \approx 
\frac{\omega^2 - \rho \hat f(0) + B + i \rho\hat f^2(0) \omega/(4\pi
c^3)} {2\rho A} \times \left( 1 - \sqrt{ 1 - 
\frac{4\rho A}{[\omega^2 - \rho \hat f(0) + B + i \rho\hat f^2(0)
\omega/(4\pi c^3)]^2}} \right) .
\label{SMALLOMEGA}
\end{equation}
We have two limiting cases. At high densities and low frequencies
($\rho \hat f(0) \gg \omega^2, B, 2\sqrt{\rho A}$), i.e.\ when the
semicircular part of the DOS does not reach low frequencies, the
square root can be Taylor-expanded, and one gets
\begin{equation}
g(\omega) \approx \frac{\omega^2}{2\pi \rho c^3},
\end{equation}
which is precisely Debye's law. At small densities, on the other hand,
the center of the semicircle (which is at $\omega^2=\rho \hat f(0) -
B$) starts to be comparable to its radius ($\propto \sqrt{\rho}$),
meaning that the states in the semicircle hybridize with the sound
waves. 
Mathematically, the instability arises when ${\cal G}(0)$ develops an 
imaginary part. This can only come from the square root in 
Eq.~(\ref{SMALLOMEGA}), and it will happen for $\rho<\rhoc$, with 
$\rhoc$ fixed by the condition
\begin{equation}
2 \sqrt{A\rhoc} + B = \rhoc \hat f(0).
\end{equation}
Now when $\rho \gtrsim \rhoc$ and $\omega \ll \omega^* = 2\pi
c^3\sqrt{\rhoc A}/(\rhoc \hat f^2(0))$, the square root in
Eq.~(\ref{SMALLOMEGA}) behaves as
\begin{equation}
\sqrt{ D(\rho-\rhoc) - i \omega/\omega^* },
\label{ORDEN}
\end{equation}
with $D$ a positive constant. Here we can distinguish two regimes:
\begin{itemize}
\item When $\omega^* D (\rho-\rhoc) \ll \omega \ll \omega^*$, the
imaginary part of $\cal G$ is proportional to $\sqrt{\omega}$, and
thus the DOS is $g(\omega) \propto \omega^{3/2}$.
\item When $\omega \ll \omega^* D (\rho-\rhoc) \ll \omega^*$, we have
an imaginary part $\propto \omega$, and $g(\omega) =
\omega^2/(\omega^* \sqrt{\rhoc A D (\rho-\rhoc)})$. So the DOS is
Debye-like, but with a very large prefactor, basically unrelated to
the speed of sound.
\end{itemize}

It is therefore natural to identify $\omegaBP$ with $\omega^* D
(\rho-\rhoc)$, that can indeed be arbitrarily small upon approaching the
instability. Notice that the mechanical instability is a kind of
phase transition, for which the order parameter is $-\mathrm{Im}\,
{\cal G}(0)$. From Eq.~(\ref{ORDEN}) we see that this order parameter
behaves as $(\rhoc-\rho)^\beta$, with $\beta=1/2$, like in mean-field
theories.

Since the behavior of the propagator at high momentum does not
strongly affect the dispersion relation~\cite{Grigera01}, we do not
expect deviations from a linear dispersion relation at $\omegaBP$.
This has been checked either numerically or solving numerically the
self-consistency equation given a particular choice for the function
$f(r)$ (see the numerical results section below).

\section{An example: the nearly-Gaussian case} \label{NEARLY-GAUSSIAN}

As previously stated, the model is completely defined when $f$ and the
distribution of the $\bm{x}$ are chosen.  As discussed in
sec.~(\ref{REAL}), a good approximation taking into account their
correlations is the replacement of the spring constants with {\em
effective} ones. That amounts to taking $f(r) = g(r)v^{\prime
\prime}(r)$ instead of $v^{\prime \prime}(r)$.

\begin{figure}[h!]
\includegraphics[angle=90,width=0.7\textwidth]{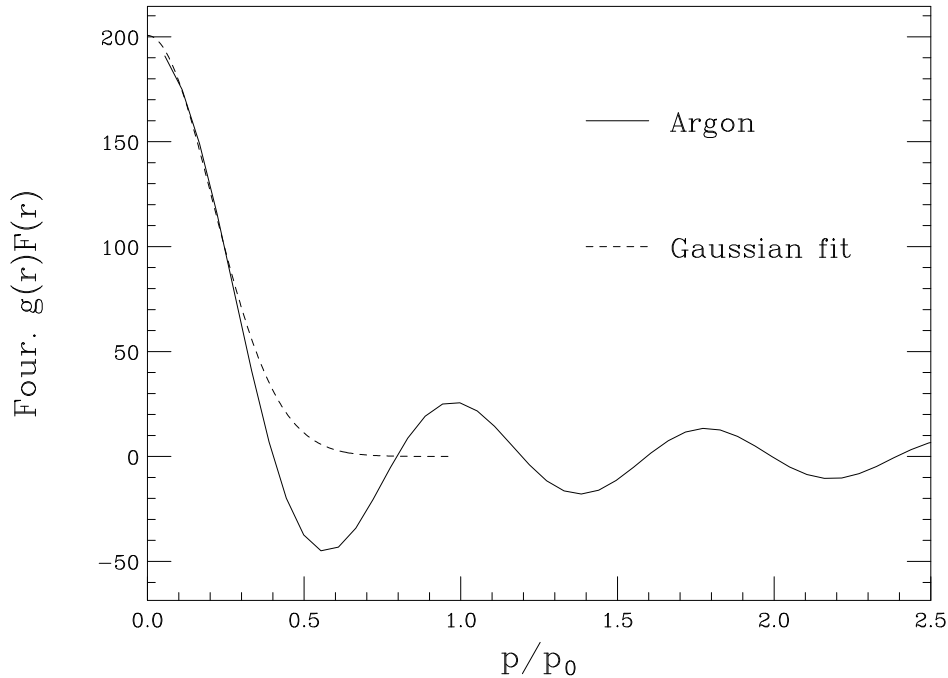}
\caption{The Gaussian choice for $f(q)$ compared with the Fourier
transform of $g(r)v^{\prime \prime}(r)$ at $\sim 10 K$, $v(r)$ being the 
Lennard-Jones pair potential which is supposed to model the Argon pair 
interactions.}
\label{CONFRONTO}
\end{figure}

As a model for the effective spring constants, we shall consider the
family of functions
\begin{equation}
f_\alpha (\bm{r})=(1-\alpha \bm{r}^2/\sigma^2)
e^{-\bm{r}^2/(2\sigma^2)},
\label{FAMIGLIA}
\end{equation}
where $0\le \alpha \le 0.2$ (the upper bound has to be imposed in
order to guarantee a positive sound velocity). When $\alpha=0$
(Gaussian case), the Hessian is strictly positive. When $\alpha>0$, we
have a stable elastic solid at high densities, while at low enough
densities, typical interparticle distances will be large enough to
allow negative eigenvalues (imaginary frequencies). Therefore the
density $\rho$ controls the appearance of a mechanical instability in
this model.  The counterpart in real glasses of that instability is
expected to be the mode-coupling transition. As a matter of fact, that
dynamical transition marks a change of the local topological
properties of the potential energy landscape~\cite{saddles}.  As a
consequence, there is a transition from a region where the short time
dynamics is ruled by a positive Hessian (high density) to one where an
extensive number of negative eigenvalues is found (low density).

The choice~(\ref{FAMIGLIA}) may seem an oversimplification, too
distant from any realistic case. However, this is not actually so, at
least for small momenta. For the sake of comparison we can see in
figure~(\ref{CONFRONTO}) the Fourier transform of the function
$g(r)v^{\prime \prime}(r)$ for Argon at very low temperature ($\sim 10
K$) together with that of~(\ref{FAMIGLIA}) in the purely Gaussian
case. Since the Fourier transform of our force decreases an order of
magnitude by $p_0=2/\sigma$, we shall take this as {\em our} $p_0$
during the following discussion, and $\sigma$ will be our unit of
length.

Choosing two different values of $\alpha=0, 0.1$, we have both
computed the DOS of this model numerically using the method of moments
\cite{Benoit92} with a box of side $L=128\,\sigma$ (more than $5\times
10^5$ particles), which allows to reconstruct the spectrum up to very
low frequencies, and numerically solved equation~(\ref{CACTUS}) for
several values of $\rho$, thus obtaining the structure factor and the
density of states.

\subsection{Theory vs. Simulations: Dynamic structure factor}

\begin{figure}[htp]
\includegraphics[angle=90,width=0.69\textwidth]{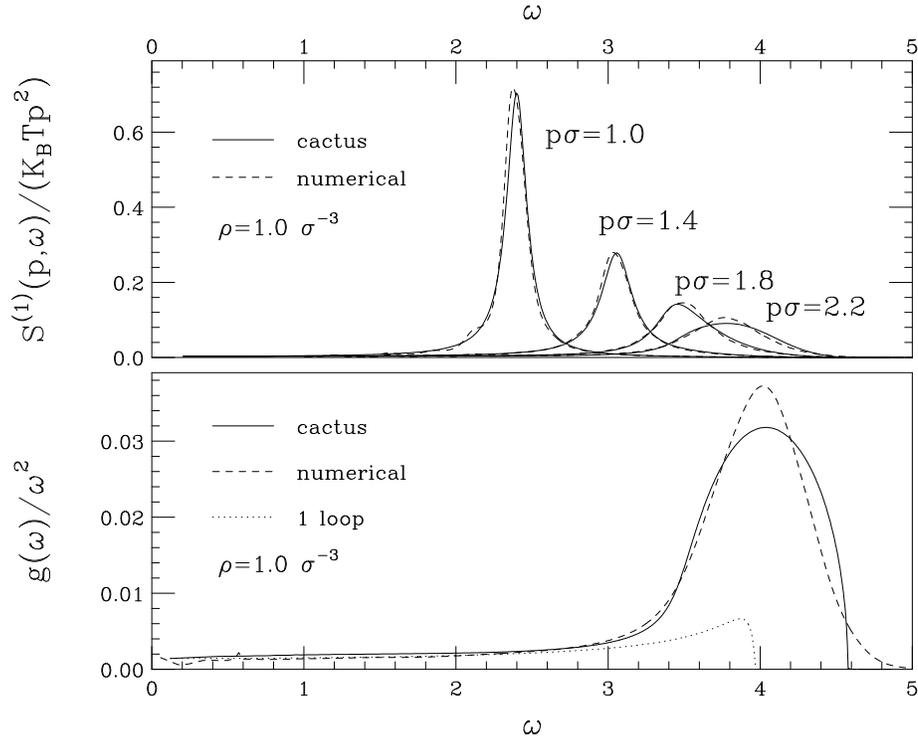}
\caption{(Top) Dynamic structure factor for different value of the
momentum, obtained both solving the integral equation derived from the
cactus resummation (theory) and using the method of moments
(simulations) in order to get the spectra of a vibrational off-lattice
model with spring constants given by~(\ref{FAMIGLIA}) with $\alpha=0$.
(For lower momenta, the comparison cannot be done due to finite volume
effects~\cite{Martin-Mayor01}).  (Bottom) The comparison between the
theoretical and experimental DOS $g(\omega)$ divided by the Debye
contribution $\omega^2$ for the same model.}
\label{comparison}
\end{figure}

In figure~(\ref{comparison}) we show both the $S(p,\omega)$ for
several values of the momentum (top) and $g(\omega)/\omega^2$ (bottom)
as obtained from equation~(\ref{CACTUS}) at $\rho=1$, together with
the results from numerical simulations for the model~(\ref{FAMIGLIA})
with $\alpha=0$.

Let us note that a very good agreement with the numerical data is
achieved.  We also found that the agreement is still satisfactorily
for densities down to $\rho \approx 0.6$. Notice that even for
$\rho=1$, the cactus resummation fails to reproduce the exponential
decay at large frequencies of the density of states.

\begin{figure}[h!]
\includegraphics[angle=90,width=0.7\textwidth]{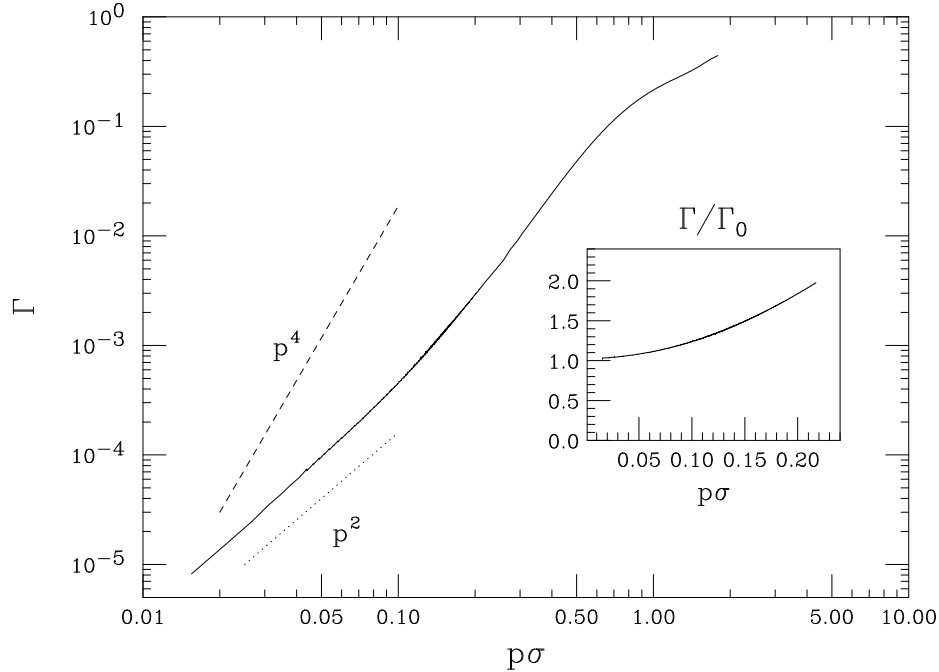}
\caption{The scaling of the broadening $\alpha$ of the Brillouin peak
for the $\alpha=0$ model as a function of the momentum $p$. (Inset)
The large q contribution $\Gamma_0$ as obtained by means of
eqs.~(\ref{GAMMA})~(\ref{sigmazero}) with respect to the total
broadening $\Gamma$.}
\label{scaling}
\end{figure}

Finally, let us look at the scaling of the width of the peak in the
frequency domain.  In the inset of figure~(\ref{scaling}) we plot
$\Gamma(p)$, obtained by means of~(\ref{GAMMA}).  As expected, the
$p^2$ scaling is found for very small momenta, which crosses over to a
region where a simple law as~(\ref{BROAD}) is not suitable to describe
the real behavior of the system.

Note that the region where the $p^2$ scaling is actually found, i.e.
$p/p_0 < 0.1$ is quite different from the region explored by X-rays
and neutrons scattering experiments, which rather spans the momenta
$0.1 < p/p_0 <0.5$. It is worthwhile to note that the same conclusion
can be drawn using MCT for hard spheres~\cite{GoMa}.

\subsection{Theory vs. Simulations: The Boson peak}

In sec.~(\ref{BOSON}) we have shown that the main result of the ERM
theory, namely the integral equation~(\ref{CACTUS}), predicts the
rising of a peak of the function $g(\omega)/\omega^2$ in the low
frequency region when at low enough densities the system approaches a
kind of phase transition where negative eigenvalues (imaginary
frequencies) begin to appear.

\begin{figure}[h!]
\includegraphics[angle=90,width=0.7\textwidth]{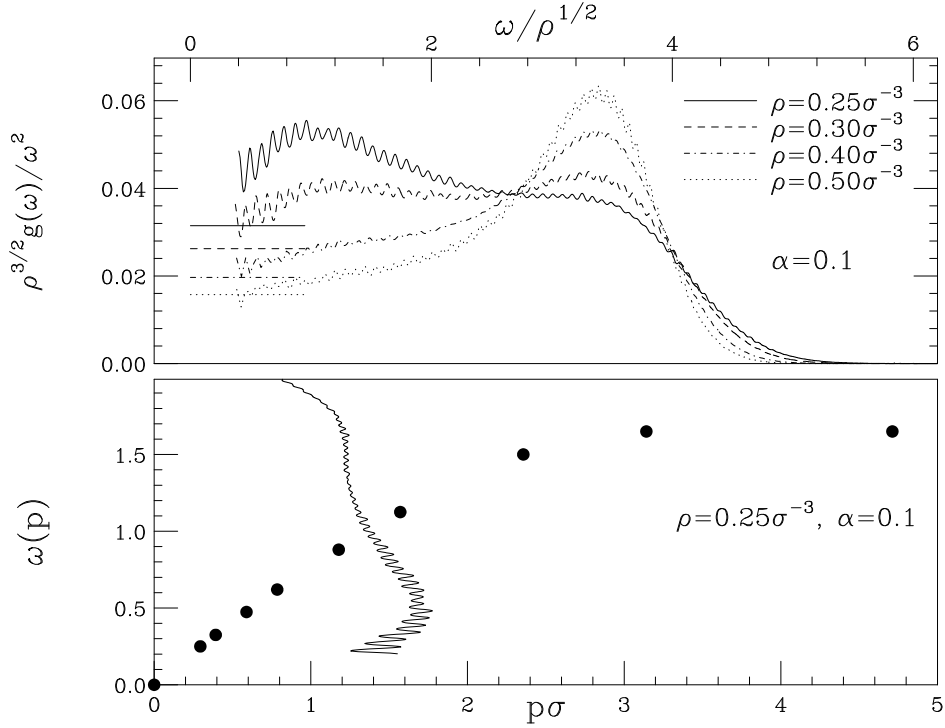}
\caption{(Top) Raising of Boson peak in the nearly-Gaussian
($\alpha=0.1$) case when the density is lowered down to the critical
density $\rho_c$.  The BP grows and shifts towards lower frequencies
upon decreasing the density.  (Bottom) $g(\omega)/\omega^2$ and
$\omega(p)$ for $\rho=0.25/\sigma^{-3}$ show that the BP lies in the
region of frequencies where the relation dispersion is still linear.}
\label{NO-GAUSSIAN-NUMERICO}
\end{figure}

This is what happens indeed in our model~(\ref{FAMIGLIA}) as shown by
the DOS obtained with the method of moments.  In
Fig.~(\ref{NO-GAUSSIAN-NUMERICO}) (top) we show the DOS divided by
$\omega^2$ at several densities for $\alpha=0.1$.  All these densities
are well above the critical density, which for this model is difficult
to locate. Anyhow, at $\rho=0.05\sigma^{-3}$ imaginary frequencies are
clearly found. As predicted, a peak in $g(\omega)/\omega^2$ arises on
approaching the instability. As density is reduced, the peak grows
(relative to the Debye value, also plotted) and moves to lower
frequencies. The maximum of the peak lies always in the linear region
of the dispersion relation (Fig.~\ref{NO-GAUSSIAN-NUMERICO},
bottom). Since decreasing density plays the role of increasing
temperature in our model, this peak reproduces the experimental
features of the BP.

\begin{figure}[h!]
\includegraphics[angle=90,width=0.6\textwidth]{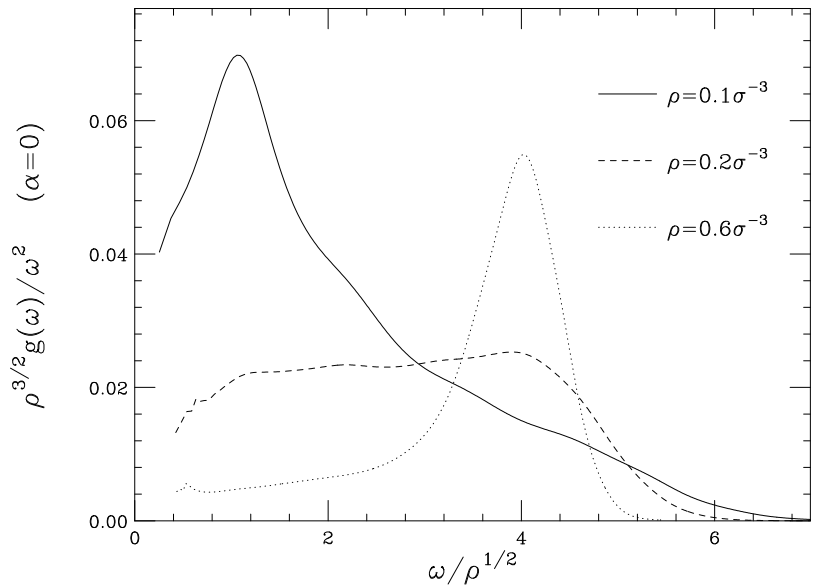}
\caption{Raising of Boson peak in the purely Gaussian ($\alpha=0$) case.}
\label{GAUSSIAN-NUMERICO}
\end{figure}

Interestingly enough, the BP even shows up in the purely Gaussian
model (see Fig.~(\ref{GAUSSIAN-NUMERICO})), which is stable at all
densities because $\alpha=0$.  This case can be thought as the
situation where the critical density is $\rhoc=0$.


Since the rising of the BP seems to be a low-density feature of
vibrational spectra and the eq.~(\ref{CACTUS}) is best suited to
describe correctly the high-density region, only a qualitative
agreement among the theory and numerics can be achieved for the
model~(\ref{FAMIGLIA}).

\begin{figure}[h!]
\includegraphics[angle=90,width=0.7\textwidth]{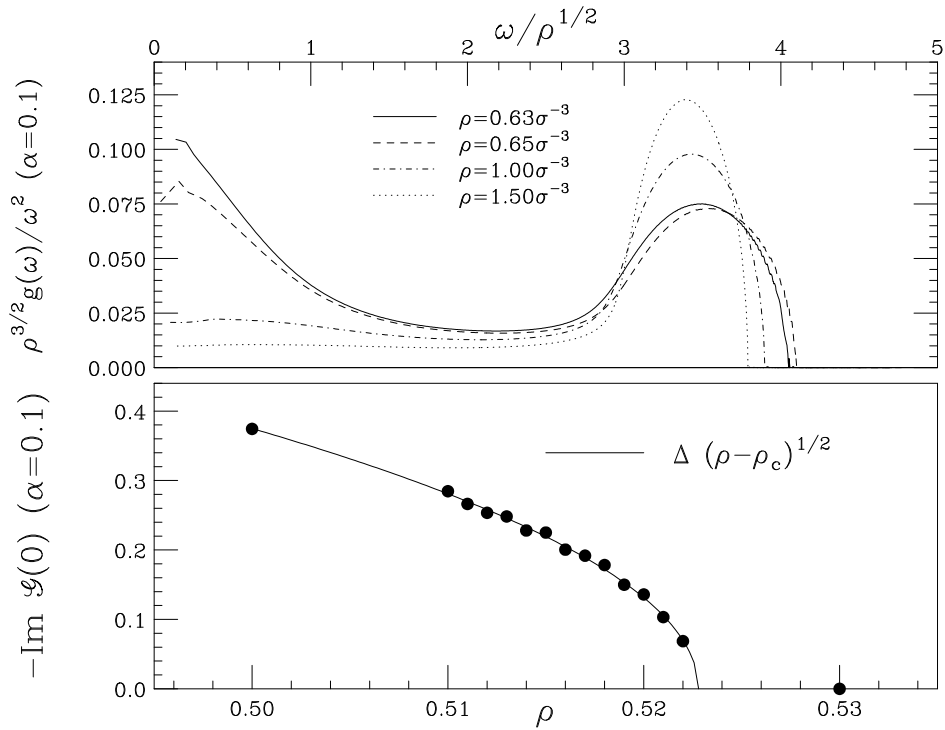}
\caption{Top: Reduced DOS divided by $\omega^2$ for $\alpha=0.1$ at
several densities, from the numerical solution of the self-consistent
equation of \protect\cite{Grigera01}. Bottom: Order parameter
($-\mathrm{Im}\, {\cal G}(0)$) of the mechanical instability phase
transition vs.\ density (points). Solid line is a fit to the predicted
behavior $\Delta (\rhoc-\rho)^{1/2}$ the fitting parameters being
$\Delta$ and $\rhoc$.}
\label{NO-GAUSSIAN-ANALITICO}
\end{figure}

This can be checked looking at Fig.~(\ref{NO-GAUSSIAN-ANALITICO}),
where we show the DOS obtained from the self-consistent $G(q,z)$, and
the behavior of $-\mathrm{Im}\, {\cal G}(0)$ with density.

\begin{figure}[h!]
\includegraphics[angle=270,width=0.55\textwidth]{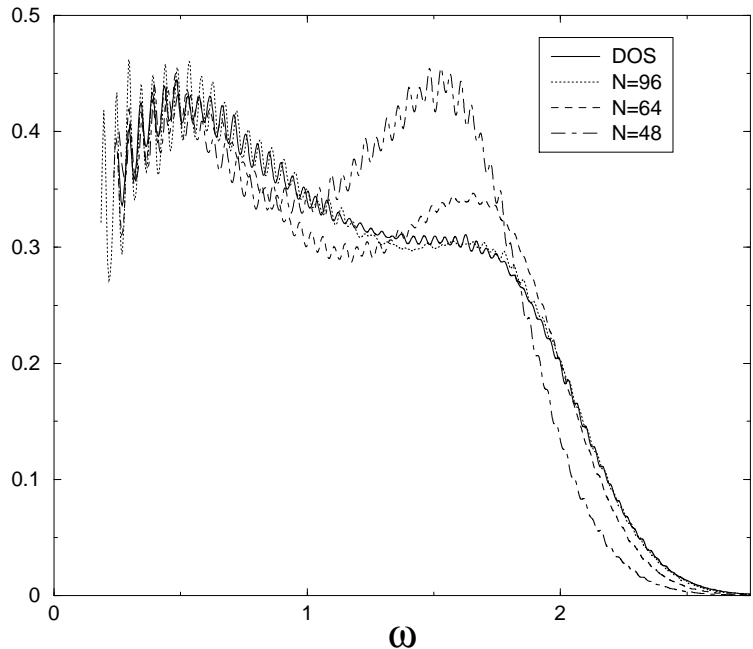}
\caption{The function $g(\omega)/\omega^2$ (labelled as DOS) and the
function $S(p,\omega)/(K_B T p^2)$ at different values of the momentum
$p$ given by $p= 2 \pi N/\sigma$ for the $\alpha=0.1$ model at
$\rho=0.25/\sigma^{-3}$.}
\label{LIMITE}
\end{figure}

As stated before, the analytical solution compares reasonably well
with the numerical solution (Fig.~\ref{NO-GAUSSIAN-NUMERICO}) on a
qualitative level, since it greatly overestimates $\rhoc$.  A still
open point is the agreement about the critical exponent $\beta$.

Finally, in the fig.~(\ref{LIMITE}) there is the numerical
confirmation the the relation~(\ref{INFINITO}) among the DOS
$g(\omega)$ and infinite momentum limit of the dynamic structure
factor $S^{(1)}(p,\omega)$ (at one-excitation level) actually
holds. Therefore we conclude that the secondary-peak identified in the
dynamic structure factor, is nothing but the Boson-Peak.

\section{Conclusions} \label{CONCLUSIONS}

In summary, we have applied the Euclidean random matrix approach to
the study of the high-frequency excitations of glassy systems.  The
main mathematical result is a non-linear integral equation whose
solution provides the dynamic structure factor (at the one excitation
approximation) and the density of states. Our approach has allowed us
to theoretically confirm, in a model-independent way, that the
Brillouin peak of the dynamic structure factor corresponds to the
propagation of sound-waves in the glass. At small momentum the width
of the Brillouin peak scales as $p^2$. At higher $p$ (actually, in the
experimentally relevant range) a more complicated law is found, but as
shown in Fig.~\ref{scaling}, the local logarithmic derivative is
always smaller than that corresponding to Rayleigh's $p^4$ law. At
still larger exchanged-momentum, the dynamic structure factor starts
to collapse onto the DOS. The frequency and width of the Brillouin
peak start to be momentum independent. The experimentally identified
secondary peak in the dynamic structure factor corresponds to the
Boson peak in an intermediate $p$ regime in which the Brillouin peak
width and position stills depends on exchanged momentum.

Our theoretical approach is also able to cope with the Boson peak. In
the family of models that we consider, there is a mechanical
instability transition controlled by the density, as signalled by the
presence of imaginary frequencies. The vibrational density of states
of our models contains a BP which is the precursor of the instability
transition. The BP in our model shares the main features of the
experimental BP: it appears for frequencies in the linear part of the
dispersion relation and it shifts towards arbitrarily low frequencies
on approaching a mechanical instability.  We also reproduce
qualitatively the behavior of the silica BP when the density
changes~\cite{SILICADENSITYEXP,SILICADENSITYNUM} (the detailed theory
of the temperature evolution of the silica BP should consider its
negative thermal dilatation coefficient).  The BP is built from the
hybridization of sound waves with high frequency modes (extended but
non propagating) that get softer upon approaching the instability.
The analogous of our instability transition in nature is the
topological phase transition~\cite{saddles} that underlies the dynamic
crossover at the Mode Coupling temperature of real
glasses~\cite{TDYN}.  The precise nature of the high-frequency modes
that hybridize with the sound-waves is most likely material dependent
and non-universal: they could be
transverse~\cite{Horbach01,Sampoli97,Ruocco00,Matic01},
optical~\cite{Taraskin99}, or even longitudinal modes as in our
model. We believe however that the basic mechanism for the formation
of the BP uncovered in our model is common to most (if not all)
structural glasses. Yet all real glasses do have transverse
excitations, and one could ask about generic new features introduced
by these modes.  Work is currently in progress to extend the ERM
approach to include transverse displacements.

\end{document}